\documentclass[11pt,amsmath,nofootinbib,axodraw]{revtex4}
\usepackage{setspace}
\usepackage{graphicx}
\usepackage{bm}
\usepackage{epsfig}
\usepackage{color}
\usepackage{colordvi}

\newcommand{\beq}{\begin{equation}}
\newcommand{\eeq}{\end{equation}}
\newcommand{\bea}{\begin{eqnarray}}
\newcommand{\eea}{\end{eqnarray}}

\newcommand{\al}{\alpha}

\newcommand{\del}{\delta}

\newcommand{\om}{\omega}

\newcommand{\mn}{{\mu\nu}}

\newcommand{\gmn}{g_{\mu\nu}}
\newcommand{\pt}{\partial}

\newcommand{\rar}{\rightarrow}

\newcommand{\intx}{\int d^{4}x}

\newcommand{\Lam}{\Lambda}

\begin{document}
\title {Reasoning by analogy: attempts to solve the cosmological constant paradox }
\author{Rafael A. Porto and  A. Zee}
\affiliation{Kavli Institute for Theoretical Physics, University of California, Santa Barbara, CA 93106}
\affiliation{Department of Physics, University of California, Santa Barbara, CA 93106}

\begin{abstract}
Talk given by one of us (A. Zee) at Murray Gell-Mann's $80^{\rm th}$ Birthday Celebration held in Singapore, February 2010. Based on R. Porto and A. Zee,  Class.\ Quant.\ Grav.\  {\bf 27}, 065006 (2010) [arXiv:0910.3716 [hep-th]]
\end{abstract}

\maketitle

The cosmological constant paradox could be summarized as follows. Expected value: $\Lam \sim	m^{4} = m/(1/m)^{3}$, an enormous energy density even if we take $m$ to be the electron mass, let alone Planck mass. Decreed value: mathematically 0, but an exact symmetry was never found. Observed value: tiny $\sim (10^{-3} eV )^{4}$	but not 0. (For the purpose of this talk, we will assume that the dark energy represents the cosmological constant.)

Can we learn something arguing by analogy? The history of physics is full of examples of analogies providing a guiding light.
The story of proton decay may provide such an analogy, as one of us proposed \cite{zdirac} some 27 years ago.
Suppose that long ago, before Murray Gell-Mann thought of quarks, perhaps in another civilization in another galaxy, a young theorist decided to calculate the rate for proton decay into $e^{+}+\pi^{0}$. It would have been natural to write down the effective Lagrangian ${\cal  L} \sim f \pi{\bar e}p$ and to compare this  with ${\cal L} \sim  g \pi {\bar n} p$, thus concluding that $f \sim \al g$ with a factor of $\al$ to account for isospin breaking.

The story of the proton decay rate would then be similar to the story of the cosmological constant as follows. Expected value: enormous since a priori Nature gave us no reason to suspect that $f$ should be tiny. Decreed value: mathematically 0, with a proof by authority (let's say Wigner)â dressed up with words like ``baryon number conservation''. Observed value: We could easily imagine that the particle physicists in the other galaxy were not as unlucky as we were, and the proton was soon observed to decay, with an extremely tiny but non-zero rate.

As is often the case in physics, the solution to the proton decay paradox did not come from thinking about the mechanism for proton decay, but from hadron spectroscopy: Quarks! (Gell-Mann, Zweig, Greenberg, Han and Nambu). With quarks, the proton decays via a dimension 6 rather than dimension 4 operator in the effective Lagrangian: ${\cal L} \sim \frac{1}{M^{2} }{\bar e}qqq$ instead of $\pi{\bar e}p$.
Remarkably, promotion from dimension 4 to 6 is enough to solve the paradox. The reason is of course that the numbers 4 and 6 appear in the exponential! This simple argument, basically dimensional analysis, could be made more respectable using modern notions of renormalization group flow and scaling (developed by Gell-Mann and Low, Wilson, and others.)

Could we try the same trick and promote\cite{zyang} the dimension of the cosmological constant term to make it less relevant at large distances compared with the Einstein-Hilbert curvature term?
One problem is that both terms, $\intx \sqrt{g} R$ and $\intx \sqrt{g}$, are made of the same kind of stuff. In the proton decay story, the recognition that hadrons and leptons are distinct provided the first step. The only difference (within our present understanding of gravity) is that curvature involves derivatives while volume doesn't. (This suggests that perhaps a foam-like structure could distinguish between the two.) In the proton decay story we somehow managed to promote the proton decay term ${\cal  L} \sim f \pi{\bar e}p$  without promoting the ${\cal L}\sim g \pi {\bar n} p$ term. The ``secret" of course is that the the proton decay term metamorphosed into a term involving a Yang-Mills gauge field, with dimension staying at 4.\\

An interesting step was taken by Polyakov who considered (in a paper titled ``Beyond spacetime" \cite{poly}) a conformally flat universe (in Euclidean signature) $\gmn=\phi(x)^{2}\del_{\mn}$. Plugging this into the Einstein-Hilbert action, one gets (in  units with $G_N=1$) the action $S ~ \intx ~[(\pt \phi)^{2}+\Lambda \phi^{4}]$ after the unwanted minus sign is removed via analytic continuation $\phi \rar i\phi$ (Lorentz group is non-compact). 

Interestingly, the cosmological term has been promoted to dimension 4 and so becomes marginal. It would be screened in the IR, but only logarithmically. Logarithmic running is not enough to bring the huge cosmological constant down to its observed value. Another serious objection is that gravity doesn't have a scalar mode! Thus the effect that Polyakov looked at appears to be purely a gauge artifact.\\

Einstein saidâ ``Physics should be as simple as possible, but not any simpler." To this we sayâ ``The solution to the cosmological constant paradox should be as crazy as possible, but not any crazier."

Let us think out of the box, but not stray too far away from it. In a recent paper \cite{portozee}, we speculated that  gravity departs from general relativity at cosmological distance scales (in what we called the ``extreme ultra infra-red" EuIR regime.) Note that work in quantum gravity and string theory has focussed almost exclusively on short distance thus far, while our proposal is in the opposite regime.\\

In condensed matter physics we are used to systems scaling differently in space and time: 
$t\rar b^z t,~~ x \rar bx$. Lorentz invariant would tell us that  $z=1$.
Here we speculate that in the EuIR gravity breaks Lorentz Invariance and scales with a dynamical critical exponent $z$ not equal to one in Fourier space:
$\om \rar b^{1/z}\om, ~~ k \rar \frac{1}{b}k$. We exploit a possible dynamical critical behavior of gravity in the EuIR regime to scale the cosmological constant. 
Our proposal is highly speculative but not outrageous.

After some simple manipulations, we find that the cosmological constant scales like $\Lambda({\rm EuIR}) \sim \Lambda({\rm IR})(\frac{l_{\rm IR}}{l_{\rm EuIR}})^{z-1}
$. (Notice that for $z=1$ we recover Polyakov's logarithmic scaling.) With $l_{\rm EuIR} \sim 10^{4} {\rm Mpc}$ the EuIR scale of the visible universe, and $l_{\rm IR}\sim 1 - 10^{3} {\rm kpc}$ the galactic or cluster scale, we have $l_{\rm EuIR} / l_{\rm IR} \sim 10^{4} - 10^{7}$. To screen the cosmological constant to the desirable value, we need $z_{{\rm EuIR}} \sim 20 - 30$, within an order magnitude of unity and at least not outrageously large. 
 By splitting spacetime into space and time we have managed to make the cosmological constant irrelevant!

Many difficulties (see our paper for a long and tortured discussion \cite{portozee}) remain. For example, we spoke of each local region of the universe trying to expand and pressing against each other in ``rebellious symphony" perhaps something like a cluster of soap bubbles.
An intriguing feature is that our action is non-local in time at cosmological distances. Perhaps a die-hard optimist could think that this could provide a hint about the nature of time.
Of course, we do not have a theory, only a speculative proposal. By splitting spacetime into space and time at (almost inconceivably) vast distances and durations, in a regime with which physics has little direct experience, we see a glimmer of a hope of understanding the cosmological constant problem.\\

We end with another possibly relevant historical analogy regarding the inverse light speed $\zeta \equiv c^{-1}$.
Consider the expected value before it was measured, again in some civilization in a galaxy far far away. The expected value is enormous in `natural units', if propagation in the ether is assumed to be similar to say sound waves in ordinary materials, let alone ocean waves (in other words, by the ``naturalness" dogma, we might have expected $\zeta$ to be comparable to $\zeta_{\rm sound}$ and we would have been off by some 6 orders of magnitude.) On the contrary, the decree (proof by authority) is that $\zeta$ is mathematically 0. Finally, it was observed by the extra-galactic version of R\/omer: the observed value turned out to be tiny but not 0 (as both Galileo and Newton thought.)

How was this $\zeta$ paradox resolved? It was resolved by making $c$ part of the kinematics. We went from the Galilean to the Lorentz group, and $c$ became a ``conversion factor" between space and time.
The unification of spacetime allows us to chose units in which $c=1$, a value protected by Lorentz invariance. In other words, it does not get renormalized! (In contrast, in non-relativistic theories $c$ would get renormalized.) Quantum fluctuations do not affect $\zeta \equiv c^{-1}$ thanks to its being part of an algebra.\\

Does this analogy tell us anything? To solve the $\zeta$ paradox, we had to go from the  Galilean group to the Lorentz group. Perhaps we need to go one step farther and extend the Lorentz group to the de Sitter group! The cosmological constant $\Lambda$, like $c$ before it, would then become a fundamental constant of nature. Just as $c$ is a fixed constant in the Lorentz algebra, $\Lambda$ then becomes a fixed constant in the de Sitter algebra. In this sense, the question why the cosmological constant is so small compared to what the ``naturalness dogma" would lead us to expect might eventually turn out to be the wrong question to ask, or at least the wrong way of phrasing the question. (We then have to ``explainâ" why the Planck mass is so large, but at least `vacuum fluctuations' would be absorbed into an overall wave-function renormalization.)\\

We have told this story happily over three birthday parties: Dirac's $80^{\rm th}$ in Coral Gables \cite{zdirac}, Yang's $85^{\rm th}$ in Singapore \cite{zyang}, and now Gell-Mann's $80^{\rm th}$ in Singapore \cite{zmgm}.\\

Happy Birthday, Murray! \\

{\it Acknowledgements:} During the course of our speculations we have benefited from enlightening conversations with Nima Arkani-Hamed, David Berenstein, Raphael Flauger, Sean Hartnoll, Jo\~ao Penedones and Joe Polchinski. We are supported in part by NSF under Grant No. 04-56556.


\begin{thebibliography}{1}

\bibitem{zdirac} A. Zee, `Remarks on the
Cosmological Constant Paradox,'' in High Energy Physics in Honor of P. A. M.
Dirac in his Eightieth Year, edited by S. L. Mintz and A. Perlmutter,
Proceedings of the 20th Orbis Scientiae (1983), p. 211, Plenum Press, New
York. 

\bibitem{zyang} A. Zee, `Gravity and its Mysteries,' in Proceedings of the Conference in Honor of C. N. Yang's $85^{\rm th}$ Birthday, edited by M.-L. Ge, C. H. Oh, and K. K. Phua, page 131, World Scientific. Also, A. Zee, Int. J. Mod. Phys. A, 23, page 1295, (2008).

\bibitem{poly} A.~M.~Polyakov, `Beyond space-time,'
  arXiv:hep-th/0602011.
  
\bibitem{portozee}  R.~A.~Porto and A.~Zee,
 `Relaxing the cosmological constant in the extreme ultra-infrared,'
  Class.\ Quant.\ Grav.\  {\bf 27}, 065006 (2010)
  [arXiv:0910.3716 [hep-th]].
  
  \bibitem{zmgm} The present volume; Proceedings of the Conference in Honor of Murray Gell-Mann's $80^{\rm th}$ Birthday, World Scientific.
  

\end{thebibliography}
\end{document}